\documentclass{IEEEmce}

\usepackage{booktabs}
\usepackage{ragged2e}
\usepackage[colorlinks,urlcolor=blue,linkcolor=blue,citecolor=blue]{hyperref}
\usepackage{xcolor}
\usepackage[table]{xcolor}
\usepackage{longtable}
\usepackage{booktabs} %for toprule, bottomrule in tables
\usepackage{fontawesome5}
\usepackage{graphicx}

\definecolor{tableGray}{RGB}{243, 244, 245}
\usepackage{upmath}
\usepackage{multicol}
\usepackage{multirow}
\usepackage{ragged2e}
\usepackage{tabularx}
\newcolumntype{P}[1]{>{\RaggedRight\arraybackslash}p{#1}}
\newcolumntype{Y}{>{\RaggedRight\arraybackslash}X}

\jvol{XX}
\jnum{XX}
\paper{XX}
\jmonth{xxx/xxx}
\publisheddate{DD MM YYYY}
\currentdate{DD MM YYYY}
\jname{IEEE Software}
\pubyear{2025}
\doiinfo{MCE.YYYY.Doi Number}

\begin{document}

%%%%
% NOTE: Add the Running Head Title below!
%%%%
% \sptitle{Empathy in Action} 

%%%%
% NOTE: The article title must be at most four lines.
%%%%
% \title{Empathy Guidelines for Real-World Software Engineering}
\title{Empathy Guidelines for Improving Practitioner Well-being \& Software Engineering Practices}
% \title{Empathy in Action: Developing and Implementing Empathy Guidelines in Real-World Software Engineering}
% \title{From Principles to Practice: Implementing Empathy Guidelines in Real-World Software Engineering}

% \title{Empathy in Action: Applying and Adapting Empathy Guidelines in Real-World Software Engineering}

%%%%
% NOTE: Authors from the same institution in the following sequence must be listed on a single line! Do NOT indicate the Corresponding Author behind the authors' name. Any acknowledgment must go after the conclusion section.
%%%%

\author{Hashini Gunatilake, John Grundy,\\ Rashina Hoda}

%%%%
% NOTE: Indicate only the university/institution without departments/subdivision, without address, without country, without IEEE membership, or without email. Do not use abbreviations.
%%%%

\affil{Monash University, Australia}

\author{Ingo Mueller}
\affil{Monash Centre for Health Research and Implementation, Monash Health, Australia}

%%%%
% NOTE: Add the Running Head and Article Titles below.
%%%%

\markboth{Empathy Guidelines for Real-World SE}{Empathy Guidelines for Improving Practitioner Well-being \& Software Engineering Practices}

% \markboth{Implementing Empathy Guidelines in Real-World SE}{From Principles to Practice: Implementing Empathy Guidelines in Real-World Software Engineering}

%%%%
% NOTE: Do not break the line after \begin{abstract} command.
%%%%

%126 words out of 150
\begin{abstract}Empathy is a powerful yet often overlooked element in software engineering (SE), supporting better teamwork, smoother communication, and effective decision-making. %In our previous study, we identified a range of practitioner strategies for fostering empathy in SE contexts. Building on these insights, 
This paper introduces 17 actionable empathy guidelines designed to support practitioners, teams, and organisations. We also explore how these guidelines can be implemented in practice by examining real-world applications, challenges, and strategies to overcome them shared by software practitioners. To support adoption, we present a visual prioritisation framework that categorises the guidelines based on perceived importance, ease of implementation, and willingness to adopt. The findings offer practical and flexible suggestions for integrating empathy into everyday SE work, helping teams move from principles to sustainable action.\end{abstract}

\maketitle

\enlargethispage{10pt}

\chapterinitial{Human aspects} play a significant role in software engineering (SE), shaping the interactions among software practitioners \cite{grundy2021addressing, gunatilake2024SLR}. One such human aspect is empathy, defined as ``the ability to experience the affective and cognitive states of another person while maintaining a distinct sense of self'' \cite{guthridge2021taxonomy}. Empathy has been shown to support both the well-being of practitioners and the effectiveness of software development practices \cite{gunatilake2025theory, gunatilake2024enablers, cerqueira2025empathy}. While empathy has been widely studied in fields such as medicine \cite{hojat2018jefferson, yu2009evaluation}, engineering \cite{hess2016voices, strobel2013empathy}, and  education \cite{blanco2017deconstructing, levy2018educating}, it remains an under-explored topic in the context of SE \cite{gunatilake2023empathy}.

% Through interviews with 22 software practitioners \cite{gunatilake2025theory}, we identified both the benefits of empathy and the consequences of its absence. Empathy improved mental health, job satisfaction, collaboration, and technical outcomes such as code quality and project success. Its absence was linked to burnout, poor team cohesion, reduced productivity, higher turnover, declining code quality, and even the potential decline of the business.
% Practitioners also shared strategies for fostering empathy or addressing empathy-related challenges. Based on these insights, we developed 17 actionable guidelines to help individuals, teams, and organisations embed empathy in practice (Table \ref{tab:Empathy Guidelines}). To develop these guidelines, we analysed the strategies from our earlier study \cite{gunatilake2025theory} using the following criteria:
\textcolor{black}{Through interviews with 22 software practitioners \cite{gunatilake2025theory}, we identified both the benefits of empathy and the consequences of its absence. Empathy was linked to improved mental health, job satisfaction, collaboration, and technical outcomes such as code quality and project success. In contrast, its absence was associated with burnout, poor team cohesion, reduced productivity, higher turnover, declining code quality, and even potential business decline.
In this same study, practitioners also shared strategies for fostering empathy and addressing empathy-related challenges. Based on these insights, we developed 17 actionable guidelines to help individuals, teams, and organisations embed empathy in practice (Table \ref{tab:Empathy Guidelines}) using the following criteria:}
\begin{itemize}
    \item \textit{Participant support}: We prioritised strategies that were frequently and consistently endorsed by participants, as these reflected strong practitioner backing.

    % \item \textit{Significance of causes of lack of empathy:} Strategies targeting widely reported or highly prevalent barriers to empathy were prioritised for their potential broad applicability and impact.

    \item \textit{Prevalence of causes of lack of empathy:} We assessed how frequently each cause of a lack of empathy was reported by participants. Causes cited by a larger number of participants were considered more important, and strategies addressing these were given higher priority.
   
    \item \textit{Real-world applicability:} The strategies were analysed for their feasibility and practical implementation at different levels including practitioner, managerial, and organisational.

\end{itemize}

\begin{table*} [htbp]
    \footnotesize
    \centering
    \caption{Actionable empathy guidelines and their applicability at practitioner, managerial, and organisation levels}
    \label{tab:Empathy Guidelines}
    % \begin{tabular}{P{0.01\textwidth} P{0.03\textwidth} P{0.03\textwidth} P{0.01\textwidth} P{0.01\textwidth} P{0.01\textwidth}}
    \setlength{\aboverulesep}{0pt}
    \setlength{\belowrulesep}{0pt}
    \setlength{\extrarowheight}{.3ex}
    \resizebox{\textwidth}{!}{
    % \begin{tabular}{lllccc}
    % \begin{tabular}{P{0.01\textwidth} P{0.02\textwidth} P{0.03\textwidth} P{0.01\textwidth} P{0.01\textwidth} P{0.01\textwidth}}
    \begin{tabularx}{\textwidth}{P{0.02\textwidth} P{0.15\textwidth} Y P{0.04\textwidth} P{0.04\textwidth} P{0.04\textwidth}}
        \toprule
        \textbf{ID} & \textbf{Guidelines} & \textbf{Description} & \multicolumn{3}{c}{\textbf{Applicability in}}\\
        % \cline {4-6}
         % &&&&&&\\
         &  &  & PL & ML & OL\\
         \midrule

        \rowcolor{tableGray}
         G1 & Fostering strong relationships & Strong relationships should be built among all stakeholders to promote empathy. \textcolor{black}{Strong relationships foster trust, open communication, and mutual understanding, which are key to cultivating empathy thereby enhancing collaboration, reducing misunderstandings, and creating a shared sense of purpose.} & \faCheckSquare & \faCheckSquare & \faCheckSquare \\
         
         G2 & Bridging technical (tech) and non-tech gap & \textcolor{black}{Empathy between tech and non-tech stakeholders enhances collaboration, and improves alignment between tech and business needs, leading to better SE outcomes. This guideline encourages assisting non-tech stakeholders better understand developers' technical work, tailoring tech explanations to the audience's understanding, and integrating cross-functional team members who understand both tech and non-tech aspects.} & \faCheckSquare & \faCheckSquare & \\
         
          \rowcolor{tableGray}
          G3 & Reducing friction among stakeholder groups &  Interactions between different SE roles, such as developers and testers, can lead to friction due to their contrasting responsibilities. & \faCheckSquare & \faCheckSquare & \faCheckSquare\\

         G4 & Encouraging bi-directional communication & \textcolor{black}{Clear and effective two-way communication is essential for fostering empathy and building strong, collaborative relationships. Two-way communication establishes trust, minimises misunderstandings, and fosters mutual understanding, which are key elements in cultivating empathy.} & \faCheckSquare & \faCheckSquare & \faCheckSquare\\
       
          \rowcolor{tableGray}
          G5 & Ensuring transparency about business goals & Transparent business goals allow practitioners to better align their work with  organisational goals. &  & \faCheckSquare & \faCheckSquare\\
         
         G6 & Use an empathetic approach to improve RE & Developers noted that unclear requirements often result from stakeholders' lack of empathy. Empathy in RE process helps developers better understand user needs, improving product quality. & \faCheckSquare &  & \\
         
          \rowcolor{tableGray}
          G7 & Collaborative problem solving & \textcolor{black}{Involving both technical \& non-technical stakeholders in resolving technical issues, enhances solution quality by considering the perspectives of all parties, which is fundamental to building empathy.} & \faCheckSquare & \faCheckSquare & \\
         
         G8 & Empathy during agile ceremonies & Empathy during sprint planning supports realistic work allocation, helps address challenges \& provide support during stand-ups, \& fosters a positive team culture in retrospectives and reviews. & \faCheckSquare & \faCheckSquare & \\
         
          \rowcolor{tableGray}
          G9 & Empathetic feedback process & Empathy during feedback helps developers feel appreciated. & \faCheckSquare & \faCheckSquare & \\
         
         G10 & Creating a safe space & \textcolor{black}{Fostering a safe environment where team members can share concerns without fear of judgement promotes well-being and positive team dynamics, enhancing mutual understanding and support. Empathy plays a key role in creating and maintaining this safe space.} & \faCheckSquare & \faCheckSquare & \faCheckSquare\\
         
          \rowcolor{tableGray}
          G11 & Backup plans to manage unexpected outcomes & Personal emergencies can impact performance, so empathetic support \& backup plans are vital to minimise project disruption. &  & \faCheckSquare & \faCheckSquare\\
         
         G12 & Flexibility in handling human issues & \textcolor{black}{Participants shared experiences where a lack of empathy during crisis situations led to negative outcomes, including resignations. In contrast, empathetic support allowed team members to handle emergencies and return with renewed loyalty. Demonstrating empathy and flexibility in these situations fosters team loyalty and contributes to project success.} & \faCheckSquare & \faCheckSquare & \faCheckSquare\\
         
          \rowcolor{tableGray}
          G13 & Emphasising real-world impact of developers' work & Developers disconnected from end-users may overlook their work's impact. Understanding this impact fosters empathy. & \faCheckSquare & \faCheckSquare & \faCheckSquare\\
         
         G14 & Building an empathetic team \& company culture & Company culture and leadership strongly influence practitioners' ability to express empathy, with leaders modelling empathy encouraging others to do the same. & \faCheckSquare & \faCheckSquare & \faCheckSquare\\

         \rowcolor{tableGray}
         G15 & Empathy education \& training & Integrating empathy into SE curricula \& workplace training prepares professionals to apply empathy in daily practice. & \faCheckSquare & \faCheckSquare & \faCheckSquare\\
         
         G16 & DEI policies & Foster empathy through DEI policies that value diversity \& promote a safe, respectful, and inclusive workplace. &  & \faCheckSquare & \faCheckSquare\\
         
          \rowcolor{tableGray}
          G17 & Managing empathy fatigue & Excessive empathy can cause fatigue, so setting boundaries and prioritising self-care is essential. & \faCheckSquare & \faCheckSquare & \faCheckSquare\\

         \bottomrule
    \end{tabularx}}
    % \end{tabular}
    \begin{flushleft}
         \textit{PL: Practitioner Level, ML: Managerial Level, OL: Organisational Level, RE: Requirement Engineering, DEI: Diversity, Equity, and Inclusion}
    \end{flushleft}
\end{table*}

\begin{figure*} [htbp]
    \centering
    \includegraphics[width=\textwidth]{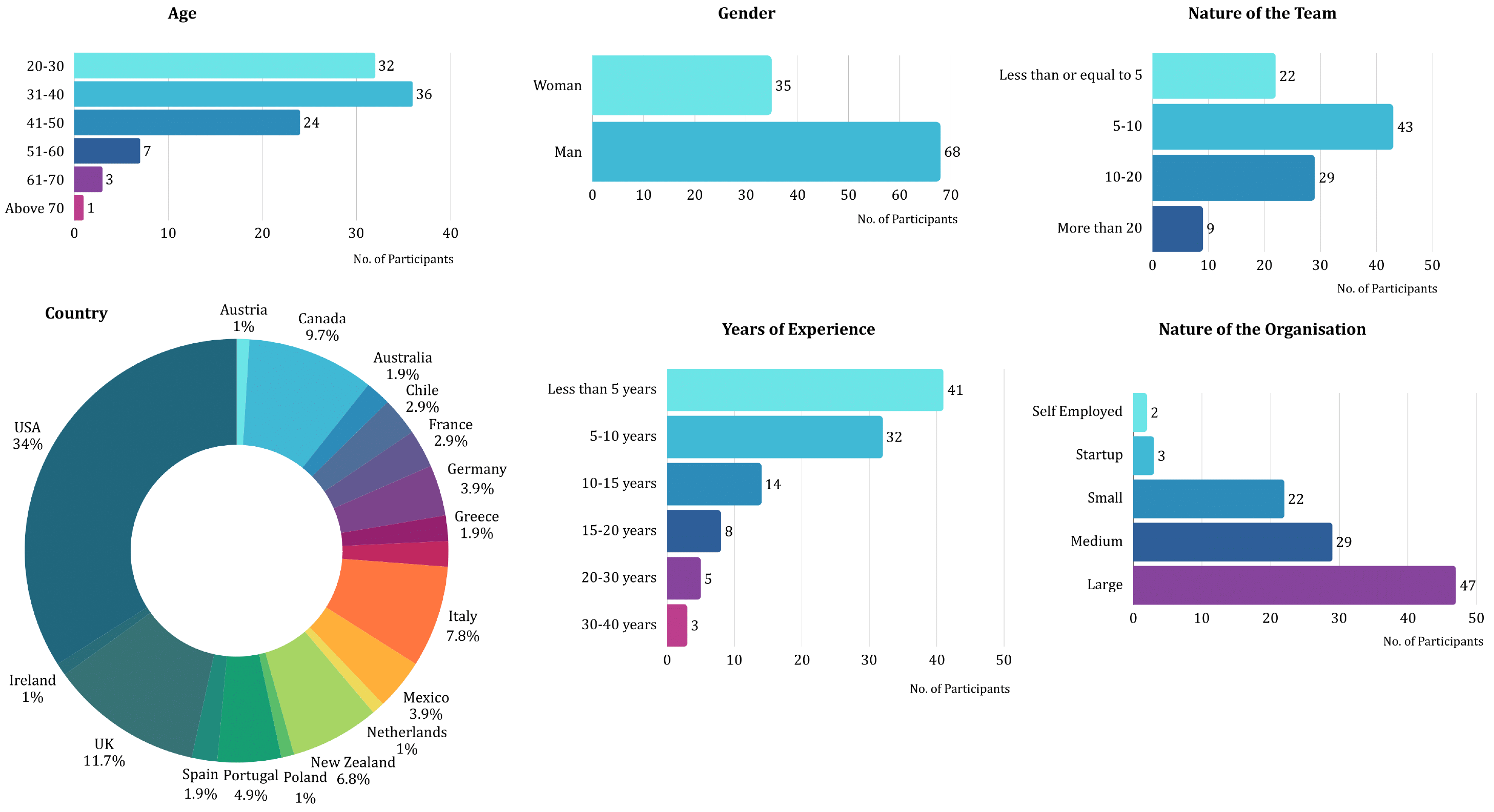}
    \caption{Overview of Demographic Information of Large-Scale Survey Participants}
    \label{fig:Demographics}
\end{figure*}

\textcolor{black}{In this paper, we introduce these guidelines for the first time and empirically evaluate them through a large-scale practitioner survey \cite{Gunatilake2025zenodo}. We conducted a survey\footnote{Approved by Monash Human Research Ethics Committee. ERM Reference Number: 45708} with software practitioners to explore the practical implementation of empathy guidelines.} %We surveyed\footnote{Approved by Monash Human Research Ethics Committee. ERM Reference Number: 45708} software practitioners to explore the practical implementation of the empathy guidelines.}
Participants were recruited via Prolific,\footnote{\url{https://www.prolific.com/}} which provided access to a diverse, globally distributed, and pre-screened pool, avoiding the sampling bias of personal or professional networks.
We applied targeted screening criteria including, participants were required to work in IT sector, have at least three years of experience, and regularly collaborate in teams. We included attention checks to ensure response quality and excluded responses that failed them. All open-ended questions were mandatory to encourage thoughtful feedback. \textcolor{black}{We collected a total of 125 survey responses. After removing responses that failed the attention-check criteria, 103 valid responses remained for analysis.} Key demographic information \textcolor{black}{is} shown in Figure \ref{fig:Demographics}.

Practitioners were asked to evaluate the importance of each guideline in real-world industry settings, assess the practicality and ease of implementation of guidelines, and indicate their willingness to adopt these guidelines in everyday SE practices. In addition, we gathered qualitative insights on how practitioners would apply these guidelines in practice, the limitations they perceived, challenges they anticipated in adoption, and suggestions for improving the guidelines to enhance their applicability. Drawing on insights from this study, this article examines how the proposed empathy guidelines can be effectively implemented in practice.
\textcolor{black}{Recent work by Cerqueira et al. proposed a conceptual framework of empathy in SE based on a grey-literature analysis, offering a complementary perspective grounded in practitioners' experiences \cite{cerqueira2025empathy}. Their study identified workplace barriers such as toxic culture and an excessive technical focus, and integrated empathy practices with effects of empathy to present a holistic model. Our work differed in that we focused on empirically evaluating the practicality and adoption of developed empathy guidelines through a large-scale practitioner survey, and on identifying challenges to their implementation along with strategies to address these challenges.} %While our focus differed, we acknowledged this complementary line of research.
\textcolor{black}{The qualitative data were analysed using thematic analysis \cite{byrne2022thematic}. We conducted an inductive coding process to identify recurring themes related to how practitioners implemented the empathy guidelines in practice, the challenges they experienced, and the strategies they proposed to address these challenges.}

%Findings

\section{Perceived Importance}
Practitioners viewed empathy as fundamental to effective SE, enabling trust, psychological safety \textcolor{black}{\cite{edmondson1999psychological}}, and communication, particularly during agile practices and stakeholder engagement. It helped bridge gaps between technical and non-technical roles, aligning goals and improving outcomes, while its absence often led to misalignment and reduced quality.
Empathy was also viewed as a motivator, connecting daily work to broader organisational goals, promoting user-centred thinking, and supporting well-being through empathetic leadership and flexible policies. Given that empathy may not come naturally to all, structured training was considered essential.
While most supported the guidelines, some warned that over-empathising could lead to fatigue and distract from core responsibilities. Overall, embedding empathy into daily practice was seen to require strong leadership, supportive policies, and ongoing training.

\begin{figure} [htbp]
    \centering
    \includegraphics[width=\columnwidth]{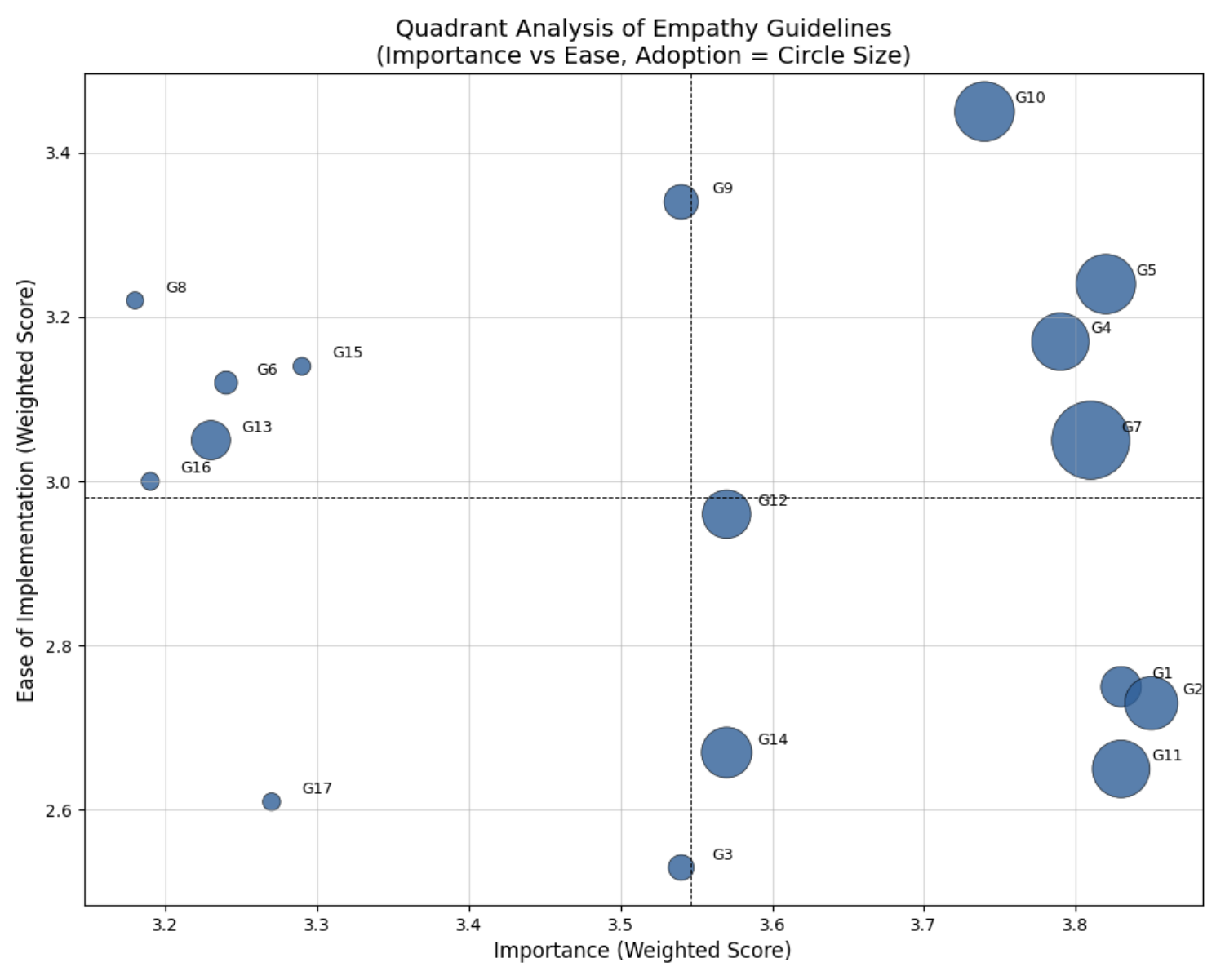}
    \caption{\textcolor{black}{A three-dimensional view of empathy guidelines' readiness: importance (x-axis), ease of implementation (y-axis), and adoption willingness (bubble size) based on practitioner ratings. Higher values on the x- and y-axes indicate higher perceived importance and greater ease of implementation, respectively. Larger bubbles indicate stronger willingness to adopt each guideline. This visualisation follows the scoring scheme used in the survey, where higher numerical values correspond to more favourable practitioner ratings, indicating greater ease rather than difficulty. All underlying numerical values used to construct the framework are available in our online Appendix(``Detailed Guidelines Analysis – Quant'' sheet)\cite{Gunatilake2025zenodo}.}} %Bubble sizes represent practitioners' willingness to adopt each guideline, with larger bubbles showing stronger adoption willingness. This chart supports implementation prioritisation.}
    \label{fig:Quadrant Analysis}
\end{figure}
% \footnotetext{https://github.com/Hashini-G/SupplementaryInfoPackage-EmpathyGuidelinesStudy}

\section{Perceived Ease of Implementation}
Practitioners noted that many empathy guidelines were relatively easy to adopt, as they aligned well with existing SE practices. Guidelines such as fostering open communication, maintaining backup plans, and applying empathy during agile ceremonies were seen as natural extensions of standard workflows. These practices typically required little cross-team coordination, making them manageable within individual teams. In smaller or more cohesive teams, integrating empathy into agile routines or offering flexible work arrangements was considered especially straightforward. Overall, ease of implementation was closely tied to how well the guidelines fit within existing team processes and organisational culture.
However, some guidelines were perceived as more difficult to apply due to structural, cultural, and resource-related challenges. These included limited managerial support for formal training, siloed communication in large organisations, and difficulty linking technical tasks to real-world impact. Tight schedules, cultural resistance, and a strong focus on deliverables over emotional engagement also hindered adoption. Sustained implementation was seen to require strong leadership and long-term commitment which were not always feasible amidst competing business demands.

\section{Adoption Willingness}
Practitioners expressed strong interest in adopting the proposed guidelines to enhance empathy in their SE practice, citing a range of compelling reasons. Many noted that the guidelines support the creation of a safe, inclusive, and collaborative work environment, which they saw as fundamental to effective teamwork and cohesion. They emphasised that empathy enables developers to better understand and respond to others' needs resulting in more inclusive solutions. They noted that fostering open communication, trust, and strong team bonds not only enhances productivity but also contributes to a healthier work culture. Several practitioners also highlighted empathy's role in reducing stress, promoting well-being, and aligning team goals, which they viewed as important for long-term performance and satisfaction. The guidelines were also recognised as consistent with growing industry attention to psychological safety and emotional intelligence, with many participants reporting that they already applied similar principles in their work. %Overall, the guidelines were considered practical and valuable for improving both project outcomes and the human experience of software development.

\begin{figure*} [htbp]
    \centering
    \includegraphics[scale=0.35]{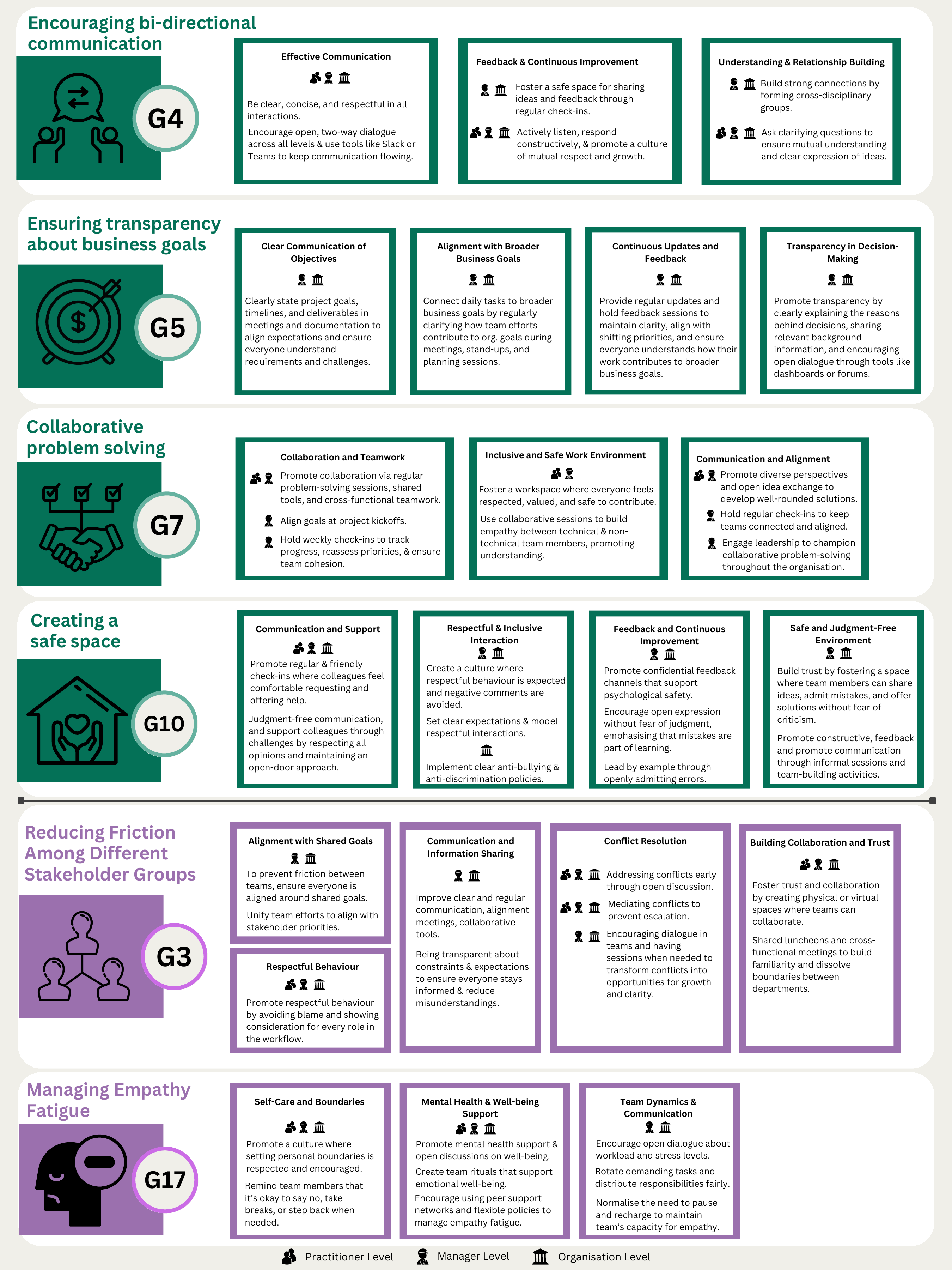}
    % \caption{Practical ways of implementing ``Quick Wins'' (High Importance, High Ease) empathy guidelines in practice.}
    \caption{Practical approaches for implementing ``Quick Wins'' empathy guidelines in green colour (high importance, high ease) and ``Lowest priority'' empathy guidelines in purple colour (low importance, low ease) at the practitioner, managerial, and organisational levels. }
    \label{fig:Quick wins and Low Priority strategies}
\end{figure*}

\section{Integrated Analysis of Importance, Ease, and Adoption Willingness}
To synthesise these views of practitioners on the proposed empathy guidelines, we created a quadrant-based bubble chart that plots importance (x-axis) against ease of implementation (y-axis), with bubble size indicating practitioner willingness to adopt each guideline. Notably, some guidelines rated as less important or harder to implement still show large bubble sizes, indicating they may gain traction if supported under the right conditions.
This visualisation enables categorisation of the guidelines into four strategic zones based on practical value and implementation feasibility, supporting prioritisation for real-world implementation.

 \begin{figure*} [htbp]
        \centering
        \includegraphics[scale=0.35]{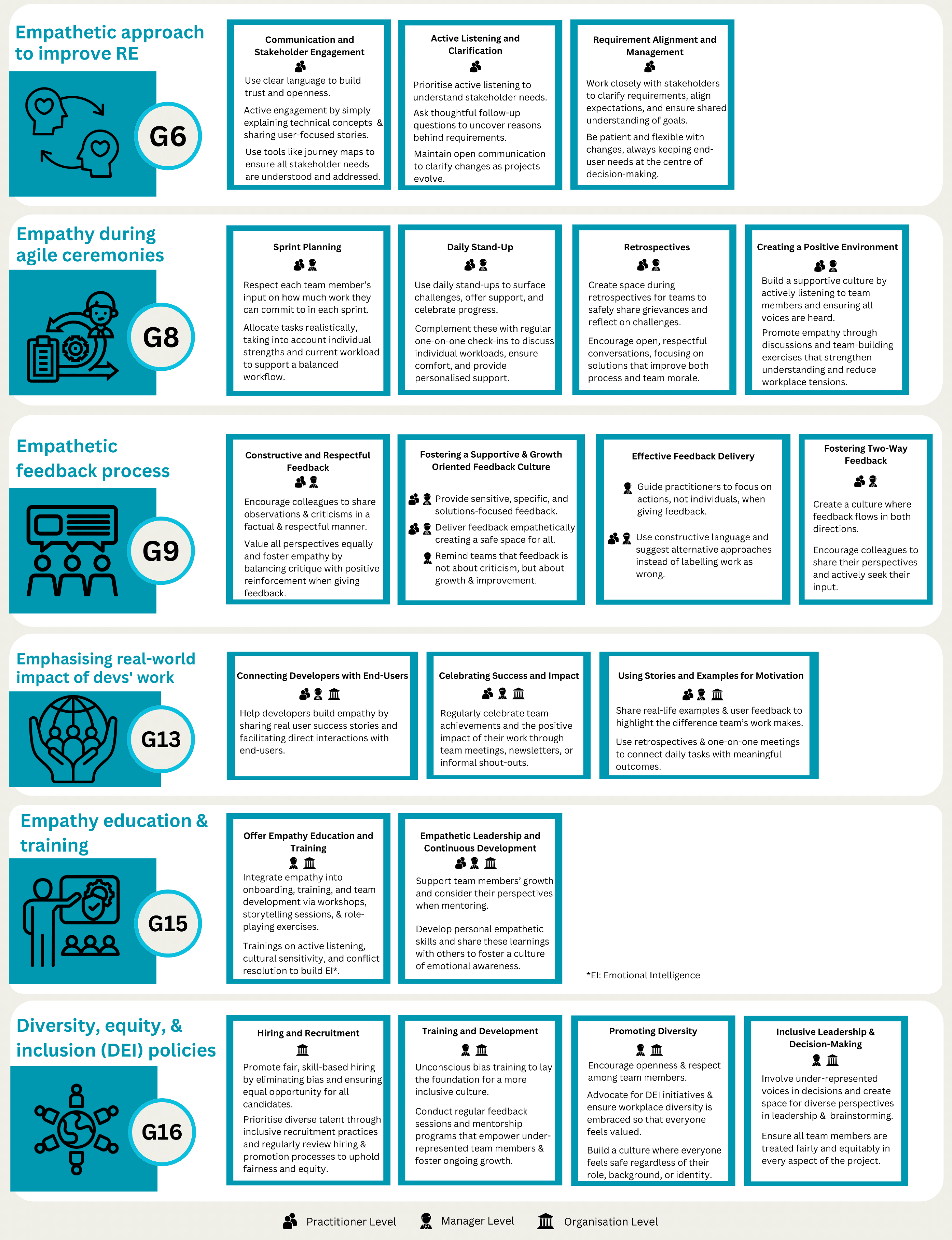}
        % \caption{Practical ways of implementing ``Nice to have'' (Low Importance, High Ease) empathy guidelines in practice}
        \caption{Practical approaches for implementing ``Nice to have'' empathy guidelines (low importance, high ease) at the practitioner, managerial, and organisational levels.}
        \label{fig:Nice to have strategies}
    \end{figure*}

    \begin{figure*} [htbp]
        \centering
        \includegraphics[scale=0.35]{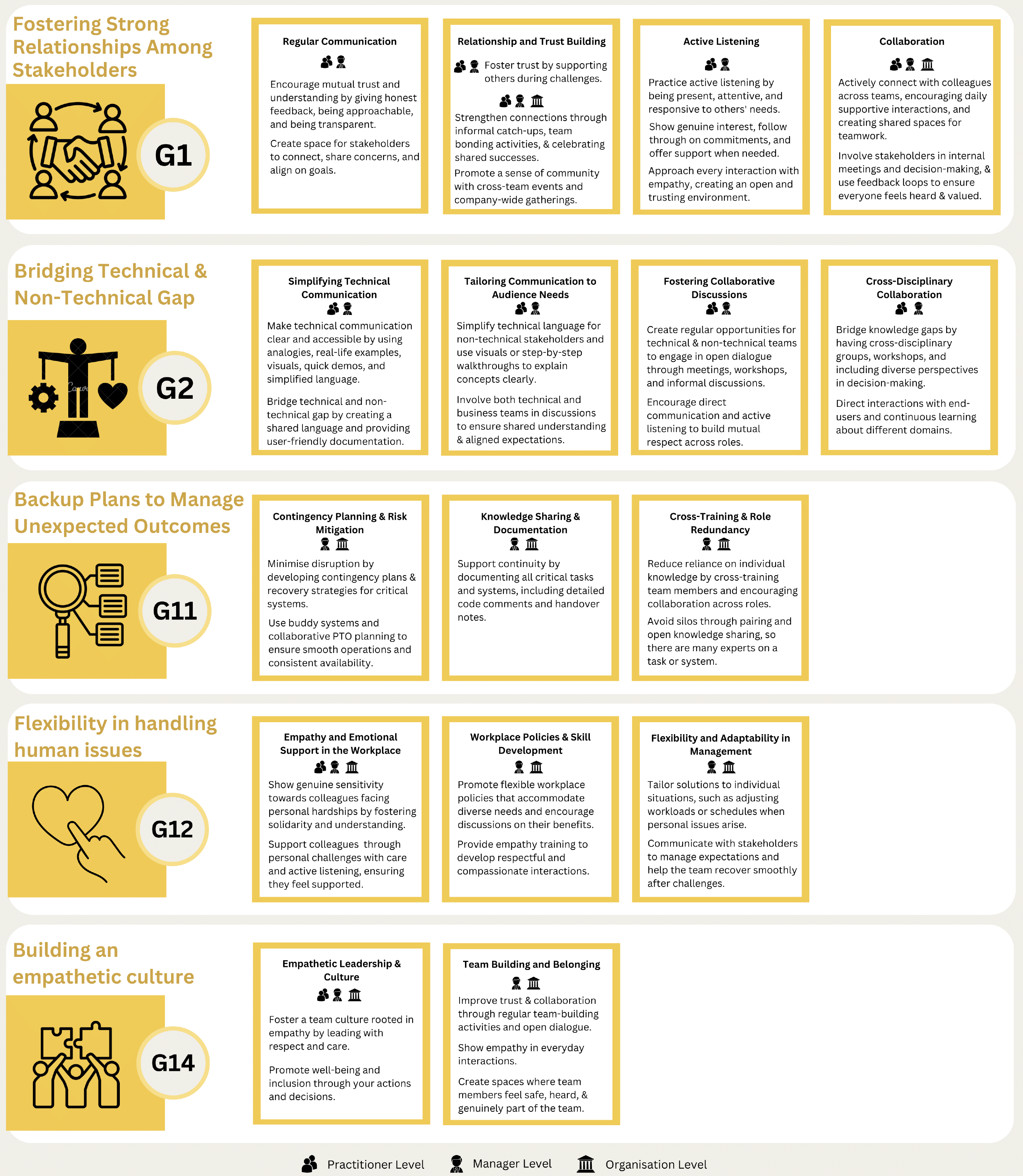}
        % \caption{Practical ways of implementing ``Strategic Investments'' (High Importance, Low Ease) empathy guidelines in practice}
        \caption{Practical approaches for implementing ``Strategic Investments'' empathy guidelines (high importance, low ease) at the practitioner, managerial, and organisational levels.}
        \label{fig:Strategic Investments strategies}
    \end{figure*}

    %   \begin{figure*} [htbp]
    %     \centering
    %     \includegraphics[scale=0.35]{Low Priority.pdf}
    %     % \caption{Practical ways of implementing ``Lowest priority'' (Low Importance, Low Ease) empathy guidelines in practice}
    %     \caption{Practical approaches for implementing ``Lowest priority'' empathy guidelines (low importance, low ease) at the practitioner, managerial, and organisational levels.}
    %     \label{fig:Lowest priority strategies}
    % \end{figure*}
    
\begin{itemize}
    \item \textbf{Quick Wins: Top Right Quadrant (High Importance, High Ease)}
    
    This quadrant includes the most promising guidelines, those that are both critical and easy to implement. Practitioners shared the practical ways of implementing our empathy guidelines in practice, as illustrated in Figure \ref{fig:Quick wins and Low Priority strategies}. Examples include G4 (Encouraging bi-directional communication), G5 (Ensuring transparency about business goals), G7 (Collaborative problem solving), and G10 (Creating a safe space). These guidelines not only received high importance ratings but were also perceived as relatively easy to adopt, with large bubble sizes reflecting strong practitioner willingness to implement them. In particular, G7 stands out with the highest adoption willingness in this group. These should be prioritised for immediate adoption in software teams.

    \item \textbf{Nice to have: Top Left Quadrant (Low Importance, High Ease)}

    Guidelines in this area, such as G8 (Empathy during agile ceremonies), were considered relatively easy to implement but were rated as less critical. These can be seen as ``low-hanging fruit'', simple enhancements that may still provide value when time and resources allow. These are low-risk, high-visibility actions that can still add value and support a more empathetic team culture.
    As illustrated in Figure \ref{fig:Nice to have strategies}, practitioners shared the practical ways of implementing these empathy guidelines in practice.

    \item \textbf{Strategic investments: Bottom Right Quadrant (High Importance, Low Ease)}

    This quadrant includes guidelines perceived as very important but difficult to implement, such as G2 (Bridging the technical and non-technical gap) and G14 (Empathetic team and company culture). These are strategic investments, which require more effort, planning, or structural support but are essential for long-term cultural change. Practitioners may need organisational buy-in or policy-level support to effectively implement these recommendations.
    Practitioners shared the practical ways of implementing these empathy guidelines in practice, as illustrated in Figure \ref{fig:Strategic Investments strategies}.

    \item \textbf{Lowest priority: Bottom Left Quadrant (Low Importance, Low Ease)}

    Guidelines here, such as G17 (Managing empathy fatigue) and G3 (Reducing friction), were rated both low in importance and difficult to implement. These are likely to be adopted more selectively or only in specific contexts where they align with team goals or organisational mandates.
    As illustrated in Figure \ref{fig:Quick wins and Low Priority strategies}, practitioners shared the practical ways of implementing these empathy guidelines in practice.

\end{itemize}

\textcolor{black}{To support practical use, the prioritisation framework can guide teams in sequencing their adoption efforts. For example, a team experiencing communication breakdowns may begin with ``quick wins'' guidelines such as G4 (Encouraging bi-directional communication) or G7 (Collaborative problem solving), which offer immediate impact with minimal resource investment. Once these foundations are in place, the team may progress to ``strategic investment'' guidelines such as G2 (Bridging the technical and non-technical gap), which require broader organisational support. This phased approach enables teams to tailor adoption to their readiness and context while gradually building a more empathetic SE environment.}

\begin{figure*} [htbp]
    \centering
    \includegraphics[scale=0.35]{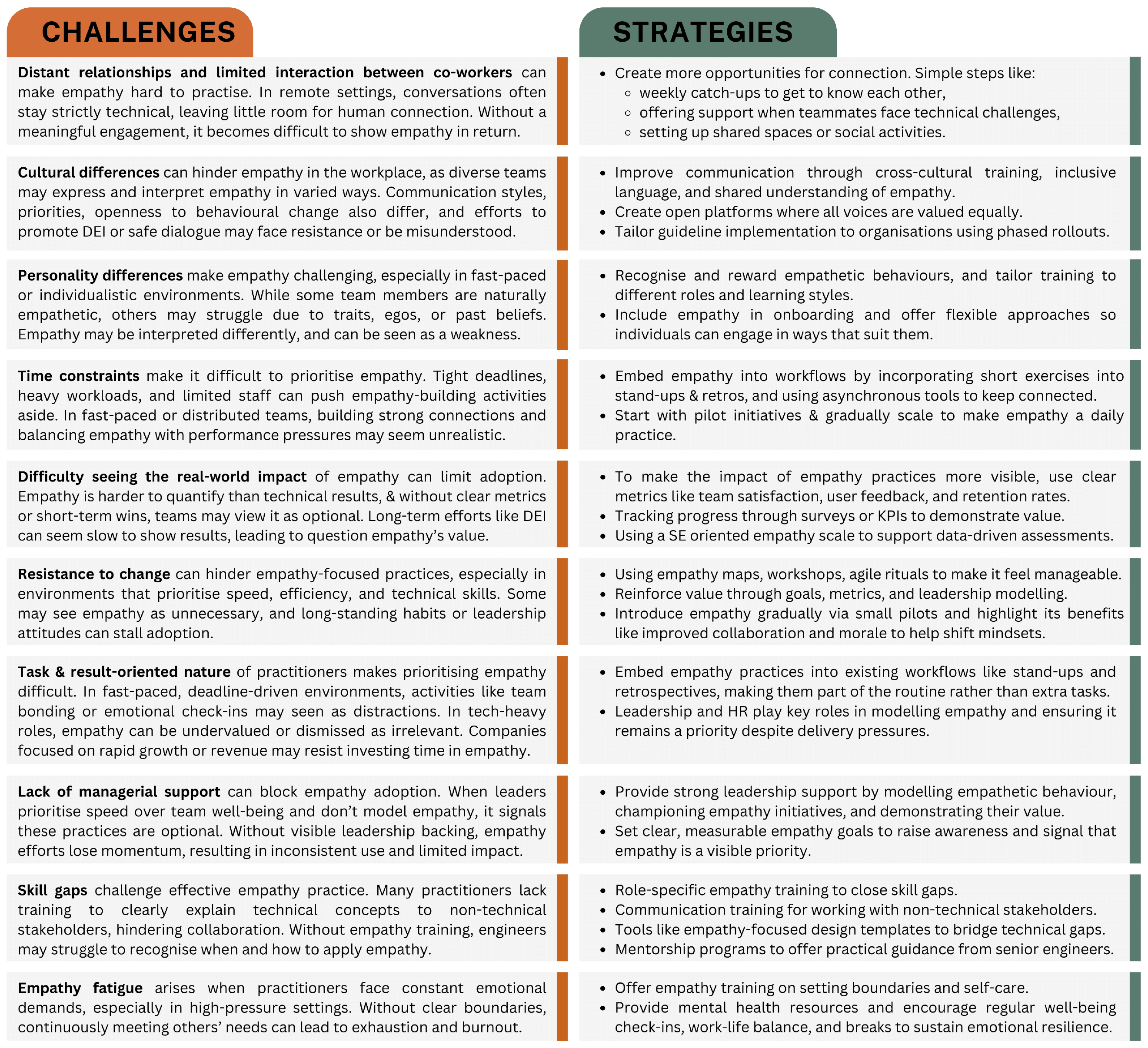}
    \caption{Challenges in Applying Empathy Guidelines and Strategies to Overcome Them}
    \label{fig:Challenges}
\end{figure*}

\section{Overcoming Adoption Challenges}
Based on practitioners' insights, we identified key challenges in adopting the proposed empathy guidelines, along with practical strategies to overcome them, as shown in Figure \ref{fig:Challenges}. \textcolor{black}{Practitioners highlighted several limitations that may hinder adoption, including time constraints, limited managerial support, and the cultural differences. In addition to these limitations, practitioners proposed a range of improvements to support successful adoption. These included integrating empathy-related practices into existing workflows, providing organisational backing through training and leadership endorsement, and promoting a culture that values open communication and psychological safety. Several participants emphasised that small, incremental changes such as regular check-ins, clearer expectations, and more intentional team interactions could make the guidelines easier to implement without imposing additional workload. Together, these challenges and strategies offer a clearer understanding of the conditions under which the guidelines are most feasible and the types of organisational and individual adjustments that could facilitate their uptake.}

\section{\textcolor{black}{Limitations}}
\textcolor{black}{Although Prolific offered global access, our sample still predominantly comprised participants from the Global North, which may have constrained the generalisability of our findings. In addition, cultural norms and organisational structures in different regions may influence how empathy guidelines are interpreted and applied. Future work should therefore recruit more participants from the Global South and examine these guidelines across varied cultural and organisational settings to strengthen representation and applicability.}

\section{Summary}
Empathy plays a powerful role in making software teams more connected, effective, and user-focused. In this study, practitioners shared how empathy helps improve communication, build stronger relationships with stakeholders, and create healthier team environments. Many found the proposed guidelines easy to adopt, especially when they aligned with existing values or everyday practices. %Others noted challenges such as time pressures, limited leadership support, and rigid organisational structures. Despite these challenges, they offered practical tips to embed empathy in daily work, through better communication, active listening, collaboration, and trust-building. 
To support real-world adoption, we also introduced an implementation prioritisation framework that helps teams decide which guidelines to focus on first. The key takeaway is that empathy is not merely a desirable addition, but a fundamental component of effective teamwork and high quality software development.

\section{ACKNOWLEDGMENT}
Gunatilake and Grundy are supported by ARC Laureate Fellowship FL190100035. We express our gratitude to all our participants for generously sharing their experiences.

\bibliographystyle{ieeetr}
\bibliography{References}

% \newpage
\begin{IEEEbiography}%[{\includegraphics[width=1in,height=1.25in,clip,keepaspectratio]{Hashini.pdf}}]
{Hashini Gunatilake} (Graduate Student Member, IEEE) is a PhD candidate at Monash University, Melbourne, Australia. She received her BSc (Hons.) in Information Systems degree from University of Colombo School of Computing (UCSC), Sri Lanka. Prior to her PhD candidature, she was in the software industry. Her research interests are software engineering, human, social \& technical aspects, human-AI interaction, agile methodology, data visualisation. More details of her research can be found at, https://hashinig.com. Contact her at hashini.gunatilake@monash.edu.
\end{IEEEbiography}

% \vspace{-1cm} 
\begin{IEEEbiography}%[{\includegraphics[width=1in,height=1.25in,clip,keepaspectratio]{John.pdf}}]
{John Grundy} (Fellow, IEEE) received the BSc (Hons), MSc, and PhD degrees in computer science from the University of Auckland, New Zealand. He is an Australian Laureate fellow and a professor of software engineering at Monash University, Melbourne, Australia. He leads the HumaniSE lab with the Faculty of Information Technology, investigating human-centric issues in SE. He is an associate editor of the IEEE Transactions on Software Engineering, the Automated Software Engineering Journal, and IEEE Software. His current interests include domain--specific visual languages, model--driven engineering, large-scale systems engineering, and software engineering education. More details about his research can be found at https://sites.google.com/site/johncgrundy/. Contact him at john.grundy@monash.edu.
\end{IEEEbiography}

% \vspace{-1cm} 
\begin{IEEEbiography}%[{\includegraphics[width=1in,height=1.25in,clip,keepaspectratio]{Rashina.pdf}}]
{Rashina Hoda} (Member, IEEE) is a Professor of Software Engineering at Monash University, Australia. Her research focuses on the human and socio-technical aspects of Software Engineering, Artificial Intelligence, and Digital Health, using qualitative and mixed methods research approaches. She was named the 2025 Top Researcher in Software Systems in Australia by The Australian. In her 2024 Springer book ``Qualitative Research with Socio-Technical Grounded Theory'', she presents a modern socio-technical variation to the Grounded Theory methods for Software Engineering and other fields. Rashina serves as an Associate Editor of the IEEE Transactions on Software Engineering, General Chair CHASE 2025, PC Co-Chair SEIP at ICSE2026, and as a member of the ICSE Steering Committee. More about her on www.rashina.com Contact her at rashina.hoda@monash.edu.
\end{IEEEbiography}

% \vspace{-1cm} 
\begin{IEEEbiography}%[{\includegraphics[width=1in,height=1.25in,clip,keepaspectratio]{Ingo.pdf}}]
{Ingo Mueller} is a project lead and user experience analyst at Monash Centre for Health Research \& Implementation and former research fellow in the HumaniSE Lab at Monash University, Melbourne, Australia. He received his PhD in computer science from Swinburne University of Technology, Australia. His research interests lie in the domain of requirements engineering including creating methods and tools that support software developers in identifying indirect stakeholders, eliciting human-centric needs and capturing and modelling these needs in close collaboration with stakeholders. Contact him at ingo.mueller@monashhealth.edu.
% More details about his research can be found at https://www.linkedin.com/in/ingo-mueller-9796088/. 
\end{IEEEbiography}

\end{document}